# Electromagnetic interpretation of leptons

## Alexander G. Kyriakos

*Saint-Petersburg State Institute of Technology, St.Petersburg, Russia*

Present address:
   *Athens, Greece*
e-mail: lelekous@otenet.gr

## Abstract

In the previous papers [1,2] on the basis of Dirac's equation we have considered the electromagnetic interpretation of the quantum theory of electron. Here we continue the electron structure study. Since the Dirac equation is also the equation of the other leptons, in present paper from the electromagnetic interpretation's point of view we analyse the structure of neutrino. We will show that not only the electron but also all other leptons have the electromagnetic structure.

PASC 12.10.-g  Unified field theories and models.
PASC 12.90.+b  Miscellaneous theoretical ideas and model.

## Contents

# 1. Introduction

In the previous paper [1] we have showed that in the presence of the particle, which has the electromagnetic field, photon is able to move along the circular trajectory and can be divided on two semi-periods, building the electron and the positron. We can suppose that the distortion origin is the electromagnetic wave gyration in the nuclear field as in the medium with high refraction index. But we have been interested mainly in the quantum mechanics interpretation, not in the features of the electron family, although they are also described by the Dirac's equation.

## 1.1 Lepton family

As it is well known, the leptons family consists of electron, heavy leptons (muon and taon), neutrinos and their antiparticles. The electron is described by Dirac's equation with a great precision. The heavy leptons description has nothing in difference with the electron, but they have greater masses. About neutrino today we only know that it doesn't have the charge, but perhaps, it has a very small mass and can oscillate between three states.

So here the following problems arise: in the frame of the electromagnetic interpretation we must

1) show how the leptons obtain the external field (i.e. the charge);

2) justify the absence of the neutrino charge and the existing of the oscillations;

3) explain the mass difference between the electron and the heavy charge leptons.

# 2. The electron external field rise

## 2.1 Gauge invariance and compensation field

The Dirac equation with the external electrical field and charge has the form [3]:

$$\left[\alpha_0(\hat{\varepsilon} - \varepsilon_{ex}) + c\hat{\vec{\alpha}}(\hat{\vec{p}} - \vec{p}_{ex}) + \hat{\beta}\, m_e c^2\right]\psi = 0\ , \qquad (2.1')$$

$$\psi^+\left[\alpha_0(\hat{\varepsilon} - \varepsilon_{ex}) - c\hat{\vec{\alpha}}(\hat{\vec{p}} - \vec{p}_{ex}) - \hat{\beta}\, m_e c^2\right] = 0\ , \qquad (2.1'')$$

where $\hat{\varepsilon} = i\hbar\dfrac{\partial}{\partial t}$, $\hat{\vec{p}} = -i\hbar\vec{\nabla}$ are the operators of energy and momentum, $\varepsilon_{ex} = e\varphi$, $\vec{p}_{ex} = \dfrac{e}{c}\vec{A}$ are the external electron energy and momentum, $c$ is the light velocity, $m_e$ is the electron mass, $\psi$ is the wave function named bispinor, $\hat{\alpha}_0, \hat{\vec{\alpha}}$ are Dirac's matrices and $\left(\varphi, \vec{A}\right)$ - 4-potential of external field.

As it is known, in modern physics the external field is added to the free electron Dirac's equation via the gauge transformation of the wave function. Thanks this transformation the additional terms, corresponding to the external electron field, appear in Dirac's equation. These terms are named the gauge field or compensation field.

Note also that in modern physics the gauge fields are associated with the transformation of the vectors, moving in the curvilinear space [4] (chapter 3).

## 2.2 Electron charge appearance

In the previous paper [1] we have obtained the free electron Dirac equations (for electron and positron):

$$\left[\left(\hat{\alpha}_o \hat{\varepsilon} + c\hat{\vec{\alpha}}\ \hat{\vec{p}}\right) + \hat{\beta}\, mc^2\right]\psi = 0 , \qquad (2.2')$$

$$\psi^+\left[\left(\hat{\alpha}_o \hat{\varepsilon} - c\hat{\vec{\alpha}}\ \hat{\vec{p}}\right) - \hat{\beta}\, mc^2\right] = 0 , \qquad (2.2'')$$

but we didn't define which mass appeared in it.

Let us remember how the mass term appears in our theory (see in detail [1]). *In accordance with our assumption, the reason for the current (mass) appearance is the electromagnetic wave motion along a curvilinear trajectory.* We showed the appearance of the current, using the general methods of the distortion field investigation [5]. For the generalisation of Dirac's equation in Riemann's geometry it is necessary [5] to replace the usual derivative $\partial_\mu \equiv \partial/\partial x_\mu$ (where $x_\mu$ is the co-ordinates in the 4-space) with the covariant derivative: $D_\mu = \partial_\mu + \Gamma_\mu$ ($\mu = 0, 1, 2, 3$ are the summing indexes), where $\Gamma_\mu$ is the analogue of Christoffel's symbols in the case of the spinors theory. When a spinor moves along the beeline, all $\Gamma_\mu = 0$, and we have a usual derivative. But if a spinor moves along the curvilinear trajectory, then not all $\Gamma_\mu$ are equal to zero and a supplementary term appears. Typically, the last one is not the derivative, but is equal to the product of the spinor itself with some coefficient $\Gamma_\mu$. Thus we can assume that the supplementary term a longitudinal field is, i.e. it is a current. So we obtain:

$$\alpha^\mu D_\mu \psi = \alpha^\mu \left(\partial_\mu + \Gamma_\mu\right)\psi , \qquad (2.3)$$

According to general theory [5] the increment in spinor $\Gamma_\mu$ has the form of the energy-momentum 4-vector. Then we have the equations:

$$\psi^+\left[\ \left(\hat{\alpha}_o \hat{\varepsilon} - c\hat{\vec{\alpha}}\ \hat{\vec{p}}\right) - \left(\hat{\alpha}_o \varepsilon - c\hat{\vec{\alpha}}\ \vec{p}\right)\ \right] = 0 , \qquad (2.4')$$

$$[ \left( \hat{\alpha}_o \hat{\varepsilon} + c\hat{\hat{\alpha}} \ \hat{\vec{p}} \right) + \left( \hat{\alpha}_o \varepsilon + c\hat{\vec{\alpha}} \ \vec{p} \right) ] \ \psi = 0 , \qquad (2.4'')$$

According to the energy conservation law we can write:

$$\hat{\alpha}_o \varepsilon \pm c\hat{\vec{\alpha}} \ \vec{p} = \mp \hat{\beta} \ mc^2 , \qquad (2.5)$$

Using (2.5) from (2.4) we obtain the usual kind of Dirac's equation with the mass (2.2).

The same result can be obtained in the vector form (see [1], Appendix 1). Let the plane-polarized photon, which has the field vectors $(E_z, H_x)$, roll up with radius $r_p$ in the plane $(x', o', y')$ of a fixed co-ordinate system $(x', y', z', o')$, so that $E_z$ is in parallel to the plane $(x', o', y')$ and $H_x$ is perpendicular to it. It could be said that the rectangular axes system $\{E_z, S_y, H_x\}$ moves along the tangent to the circumference, where $\vec{S}_y = \left[ \vec{E} \times \vec{H} \right]_y$ is the $y$-component of the Poynting vector (see Fig.1):

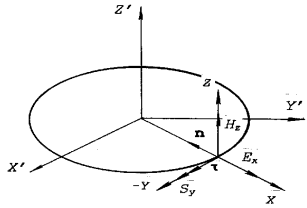

Fig.1

Let us show that due to the photon electromagnetic wave distortion the displacement ring current arises.

According to Maxwell [6] the displacement current is defined by the equation:

$$j_{dis} = \frac{1}{4\pi} \frac{\partial \vec{E}}{\partial t}, \qquad (2.6)$$

The electrical field vector $\vec{E}$, which moves along the curvilinear trajectory (let it have the direction from the centre), can be written in form:

$$\vec{E} = -E \cdot \vec{n}, \qquad (2.7)$$

where $E = \left| \vec{E} \right|$ and $\vec{n}$ is the normal unit-vector of the curve (having direction to the center) (see Fig.2):

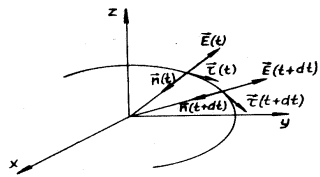

Fig.2

(Note, that further we name the spinning photon as the *gyrophoton*).

The derivative of $\vec{E}$ with respect to $t$ can be represented as:

$$\frac{\partial \vec{E}}{\partial t} = -\frac{\partial E}{\partial t} \vec{n} - E \frac{\partial \vec{n}}{\partial t} \ , \qquad (2.8)$$

Here the first term has the same direction as $\vec{E}$. The existence of the second term shows that at the wave distortion the supplementary displacement current appears. It is not difficult to show that it has a direction, tangential to the ring:

$$\frac{\partial \vec{n}}{\partial t} = -\frac{\upsilon_p}{r_p} \vec{\tau} \ , \qquad (2.9)$$

where $\vec{\tau}$ is the tangential unit-vector, $\upsilon_p \equiv c$ is the photon velocity. Then the displacement current of the ring wave can be written:

$$\vec{j}_{dis} = -\frac{1}{4\pi} \frac{\partial E}{\partial t} \vec{n} + \frac{1}{4\pi} \omega_p E \cdot \vec{\tau} \ , \qquad (2.10)$$

where $\omega_p = \dfrac{c}{r_p} = \dfrac{m_p c^2}{\hbar}$ is the angular velocity (or angular frequency),

$m_p$ is the mass of the gyrophoton; $\vec{j}_n = \dfrac{1}{4\pi} \dfrac{\partial E}{\partial t} \vec{n}$ and $\vec{j}_\tau = \dfrac{\omega_p}{4\pi} E \cdot \vec{\tau}$ are

the normal and tangent components of the gyrophoton current correspondingly. It is not difficult to see that the tangent current corresponds to the analogue of Christoffel's symbols $\Gamma_\mu$.

*Therefore, the mass that the gyrophoton equation contains, isn't the electron mass, but the gyrophoton mass and each of the above electron equations* (2.2) *contains the mass, which is equal to two electron masses:*

$$\left[ \left( \hat{\alpha}_o \hat{\varepsilon} + c\hat{\vec{\alpha}} \ \hat{\vec{p}} \right) + 2\hat{\beta} \ m_e c^2 \right] \psi = 0 \ , \qquad (2.11')$$

$$\psi^+ \left[ \left( \hat{\alpha}_o \hat{\varepsilon} - c\hat{\vec{\alpha}} \ \hat{\vec{p}} \right) - 2\hat{\beta} \ m_e c^2 \right] = 0 \ , \qquad (2.11'')$$

After the electron-positron pair production, i.e. after the gyrophoton is divided to two semi-period, the electron and positron, must go away the one from the other to become independent. But at this instant the electron and positron acquire the electromagnetic fields, and one particle moves in the field of the other (Fig.3)

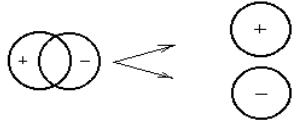

Fig.3

Therefore, the equations, which arise after the gyrophoton equation division can not be the free electron-positron equations, but the equations with the external field.

Obviously, for the particle division and remotion from one another, the energy for the electromagnetic field creation must be expended. In fact, being the particles combined, the system doesn't have any field. At very small distance they create the dipole field and at a distance, greater than the particle radius, the electron and positron acquire the full electromagnetic field. As it is known [6], the potential $V_p$ of two point charges in the point $P$ (see Fig.4)

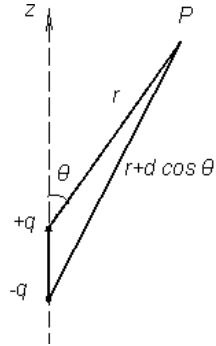

Fig.4

is defined as:
$$V_p = \frac{q}{\varepsilon_0}\left(\frac{1}{r} - \frac{1}{r+d\cos\theta}\right), \qquad (2.12)$$

where $\pm q$ is the charges, $d$ is the distance between the charges. From this formula we can obtain two limit cases:

$$\lim_{d\to 0} V_p = 0 \quad \text{and} \quad \lim_{d\to\infty} V_p = \frac{q}{\varepsilon_0}\frac{1}{r}, \qquad (2.13)$$

During the division process the particle charges appear. For the entire particle remotion the work needs to be fulfiled against the attractive forces. (In our case we can not calculate the work, using the known classical relations, since the particles are not point).

The question arises: from where does this division energy emerge? Obviously, the division energy is the field production energy. Therefore, the energy value $\varepsilon = m_e c^2$ defines the external particle field. In other words, in the paper [1] we have not obtained the free electron equation, but the equation with external field. Actually we can write the equations (2.11) in the form:

$$\left(\hat{\alpha}_0\hat{\varepsilon} + c\hat{\vec{\alpha}}\,\hat{\vec{p}} + \hat{\beta}\,m_e c^2 + \hat{\beta}\,m_e c^2\right)\psi = 0, \qquad (2.14)$$

Using the linear equation of the energy conservation, we can write:

$$\hat{\beta}\, m_e c^2 = -\varepsilon_{ex} - c\hat{\vec{\alpha}}\; \vec{p}_{ex} = -e\varphi - e\hat{\vec{\alpha}}\; \vec{A} \;, \qquad (2.15)$$

Putting (2.15) in (2.14) we obtain the Dirac equation with external field (2.1).

**From above follows:**

**Firstly** that in the initial state the gyrophoton isn't an absolutely neutral particle, but a dipole; therefore, it must have the dipole momentum.

**Secondly**, the formula (2.15) shows that in the relativistic case the mass isn't equivalent to the energy, but to the 4-vector of the energy-momentum; from this follows that energy has the kinetic origin.

**Thirdly**: let's compare the decomposition (2.15) with the energy decomposition for free electron nonlinear equation (see [1]):

$$\hat{\beta}\, m_e c^2 = -\varepsilon_{in} - c\hat{\vec{\alpha}}\; \vec{p}_{in} = -e\varphi - e\hat{\vec{\alpha}}\; \vec{A} \;, \qquad (2.16)$$

As we see, the values $(\varepsilon_{in}, \vec{p}_{in})$ describe the inner field, and the values $(\varepsilon_{ex}, \vec{p}_{ex})$ the external field of electron. When we consider the electron from great distance, the field $(\varepsilon_{in}, \vec{p}_{in})$ works as the mass, and the term $(\varepsilon_{ex}, \vec{p}_{ex})$ describes in detail the external electromagnetic field (we have linear Dirac's equation). Inside the electron the term $(\varepsilon_{in}, \vec{p}_{in})$ is needed for the detailed description of the inner field of an electron (and we have [1] non-linear Dirac's equation).

# 3. Neutral leptons (neutrino)

According to modern theory the neutrinos are described by Dirac's equation; they have the spin equal to $\frac{1}{2}\hbar$ and the charge equal to zero. Until today (STM model) the neutrino mass was supposed to be also equal to zero, but the solar oscillation experiments showed that the neutrino has a mass, although it is very small.

According to our theory the neutrino is the spinning semiphoton and must have the non zero mass. The problem appears: what is the reason why the spinning semiphoton doesn't have a charge?

## 3.1 Neutrinos production hypothesis

For the solution of the neutrino zero charge problem we suggest the following hypothesis: **neutrinos are the spinning semiphotons with circular polarization.**

Let's prove this suggestion.

If the initial photon is the plane polarized electromagnetic wave and the electrical field vector lies in the rotation plane, the vector projection on the rotation plane has one sign. In this case (see [1]) the electrical current appears; (note, that the magnetic vector projection is equal to zero and doesn't create the current).

But if the initial photon is the circular polarized electromagnetic wave, the electrical field vector doesn't lie in the rotation plane, as well as the magnetic field vector. In this case the currents have an alternate sign and integrally can give the zero value. But the performance of the other requirement is also needed here: the current alternation must have one entire period. As we know from the electromagnetic theory, the polarization vector's rotation frequency is equal to the frequency of the wave and, in fact, makes one entire revolution in one period of time. Since [1] the semiphoton length is equal to the gyrophoton length, we can suggest that the semiphoton rotation has also one entire period. In this case really the integrated electric current is equal to zero (as well as magnetic).

Now we must show that the Dirac equation in the general case contains two circular polarized waves. Consider the equations (2.2). In electromagnetic form this equations have the form:

$$\frac{1}{c}\frac{\partial \vec{E}}{\partial t} - \vec{\nabla}\times\vec{H} = -i\frac{mc}{\hbar}\vec{E}, \qquad (3.1')$$

$$\frac{1}{c}\frac{\partial \vec{H}}{\partial t} + \vec{\nabla}\times\vec{E} = -i\frac{mc}{\hbar}\vec{H}, \qquad (3.1'')$$

For the photon with the $y$ – direction we have:

$$\frac{1}{c}\frac{\partial E_x}{\partial t} - \frac{\partial H_z}{\partial y} = ikE_x \qquad\qquad \frac{1}{c}\frac{\partial E_x}{\partial t} + \frac{\partial H_z}{\partial y} = -ikE_x$$

$$\frac{1}{c}\frac{\partial E_z}{\partial t} + \frac{\partial H_x}{\partial y} = ikE_z , \qquad (3.2') \qquad \frac{1}{c}\frac{\partial E_z}{\partial t} - \frac{\partial H_x}{\partial y} = -ikE_z , \qquad (3.2'')$$

$$\frac{1}{c}\frac{\partial H_x}{\partial t} + \frac{\partial E_z}{\partial y} = -ikH_x \qquad\qquad \frac{1}{c}\frac{\partial H_x}{\partial t} - \frac{\partial E_z}{\partial y} = ikH_x$$

$$\frac{1}{c}\frac{\partial H_z}{\partial t} - \frac{\partial E_x}{\partial y} = -ikH_z \qquad\qquad \frac{1}{c}\frac{\partial H_z}{\partial t} + \frac{\partial E_x}{\partial y} = ikH_z$$

The Dirac equation solution has the harmonic wave view:

$$\psi_j = A_j e^{-\frac{i}{\hbar}(\varepsilon t - \vec{p}\vec{r})}, \qquad (3.3)$$

where the amplitudes $A_j$ are the numbers ( $j = 1,2,3,4$ ). Here we can put

$$A_j = A_0 , \qquad (3.4)$$

Since the currents don't define the poloidal rotation, we can put them equal to zero and analyse the homogeneous equations. The trigonometric form of the equation solutions are:

$$\begin{cases} E_x = A_0 \cos(\omega\, t - ky) \\ H_z = -A_0 \cos(\omega\, t - ky) \\ E_z = -A_0 \sin(\omega\, t - ky) \\ H_x = -A_0 \sin(\omega\, t - ky) \end{cases}, (3.5') \qquad \begin{cases} E_x = A_0 \cos(\omega\, t - ky) \\ H_z = A_0 \cos(\omega\, t - ky) \\ E_z = -A_0 \sin(\omega\, t - ky) \\ H_x = A_0 \sin(\omega\, t - ky) \end{cases} \quad (3.5'')$$

Consider the rotation of vectors $\vec{E}$ and $\vec{H}$ in the $XOZ$ plain. Putting $y = 0$ we obtain:

$$\vec{E} = A_0\left(\vec{i}\cos\omega\, t - \vec{k}\sin\omega\, t\right), \qquad (3.6') \qquad \vec{E} = A_0\left(\vec{i}\cos\omega\, t - \vec{k}\sin\omega\, t\right), \qquad (3.7')$$

$$\vec{H} = A_0\left(-\vec{i}\sin\omega\, t - \vec{k}\cos\omega\, t\right), \quad (3.6'') \qquad \vec{H} = A_0\left(\vec{i}\sin\omega\, t + \vec{k}\cos\omega\, t\right), \qquad (3.7'')$$

The direction of the semiphoton motion is defined by the Poynting vector:

$$\vec{S}_0 = \vec{E} \times \vec{H} = -\vec{j}\left(E_x H_z - E_z H_x\right), \qquad (3.8)$$

Calculating the above we have

$$\vec{S}_0 = A_0^2 \vec{j} \quad , \qquad (3.9)$$

and

$$\vec{S}_0 = -A_0^2 \vec{j} \quad , \qquad (3.10)$$

Thus, the photons of the right and left systems (3.5') and (3.5'') move in the contrary directions.

Fixing the vector $\vec{E}, \vec{H}$ positions in two successive time instants, the rotation direction we can define: in the initial instant ($t = 0$) and through the little time period $\Delta\, t$. We obtain the figures 5 and 6 correspondingly:

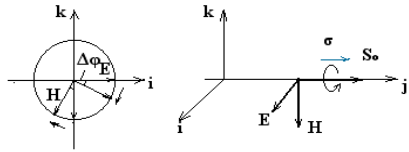

Fig. 5

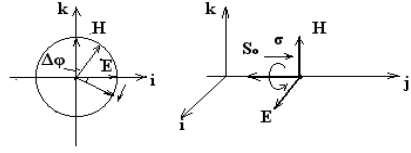

Fig.6

On the same figures we have drawn these semiphotons in motion, taking in account the motion directions.

As we see (Fig.5,6), the above semiphotons have the contrary rotations of the polarisation plains. In the first case the rotation vector and the Poynting vector have the same directions; in the second case they are contary.

In the quantum mechanics there is the value named *helicity* (see, e.g., [7]). By definition the helicity is the particle spin projection on the particle momentum direction. The helicity can be also introdused in the classical electrodynamics [8] (see chapter 8).

By same way we can determine the *inner helicity* as the projection of the poloidal rotation momentum on the momentum of the ring field motion.

As it is known (see e.g.[9]) the helicity is described by $\hat{\alpha}_5 = \hat{\alpha}_0 \hat{\alpha}_1 \hat{\alpha}_2 \hat{\alpha}_3$ matrix (note that the value $\psi^+ \hat{\alpha}_5 \psi$ according to the electromagnetic interpretation [1] the pseudoscalar $\psi^+ \hat{\alpha}_5 \psi = \vec{E} \cdot \vec{H}$ is). The contrary of the helicities of the particle-antiparticle is not difficult to show. Multiplying the Dirac equation (2.2) on $i \hat{\alpha}_5 \hat{\beta}$ and taking in account that

$$i \hat{\alpha}_5 \hat{\beta} \, \vec{\alpha} = \hat{\vec{\sigma}} \, , \qquad (3.11)$$

where $\hat{\vec{\sigma}}$ are the spin matrix, we obtain:

$$\left( i \hat{\beta} \hat{\varepsilon} - c \hat{\vec{\sigma}} \hat{\vec{p}} + imc^2 \hat{\alpha}_5 \right) \psi = 0 \, , \qquad (3.12')$$

$$\left( i \hat{\beta} \hat{\varepsilon} + c \hat{\vec{\sigma}} \hat{\vec{p}} - imc^2 \hat{\alpha}_5 \right) \psi = 0 \, , \qquad (3.12'')$$

Then

$$\hat{\alpha}_5 = \frac{c \hat{\vec{\sigma}} \; \vec{p}}{i \hat{\beta} \hat{\varepsilon} + mc^2} \, , \qquad (3.13')$$

and

$$\hat{\alpha}_5 = \frac{-c \hat{\vec{\sigma}} \; \vec{p}}{i \hat{\beta} \hat{\varepsilon} - mc^2} \, , \qquad (3.13'')$$

As it is known [7], the helicity is the Lorentz-invariant value only for the massless paticles, but it is also used for the description of the parity non-conservation of the massive particles.

From above follows that according to our theory inside the particle the operator $\hat{\alpha}_s$ doesn't discribe the particle rotation as it is preposed in the quantum mechanics, but the poloidal rotation of the fields (Fig.7)

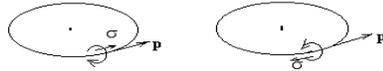

Fig.7

Let's unite the figures 5 and 6 on one drawing (Fig.8):

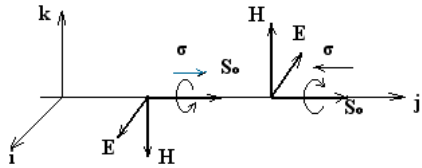

Fig.8

It is not difficult to see that both semiphotons correspond to one photon (Fig.9):

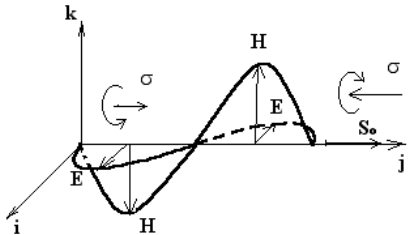

Fig.9

Let's make the **summary of this chapter**:
we suggested that *the neutrino is the spinning semiphoton which has the circular polarization;* from the above hypothesis the next conclusions follow:

**1)**as a spinning semiphoton, the neutrino has a mass;

**2)**the neutrino has the azimuthal rotation of the field, named spin, and at the same time, via circular polarization, it has a second rotation - the poloidal field rotation, named *inner helicity*.

**3)** from electromagnetic theory it follows that the angular frequency of the two rotations, mentioned above, is equal. (Note that it is not difficult classically to calculate the azimuthal rotation momentum (spin)[1], but in the classical mechanics the meaning of the poloidal rotation momentum (inner helicity) doesn't exist).

**4)** in the decay instant of the initial linear photon the neutrino and antineutrino have contrary helicities; therefore, neutrino has **only** one sign of helicity (also the antineutrino). From this fact the parity non-conservation is follows.

**5)** the helicity is linked to the second invariant of Maxwell's theory $(\psi^+ \hat{\alpha}_s \psi)^2 = (\vec{E} \cdot \vec{H})^2$.

We think the above representations allow also to calculate the mechanical end electromagnetic characteristics of neutrino.

## 3.2. About neutrino oscillations and masses

The modern investigations show that the lepton oscillation problem are the same with the hadrons. It is also known, that this problem has links with the particle mass problem. Therefore it will be right, if we will analyse these problems in a separate paper as a common case as all elementary particles mass problem.

# Appendix I. About electron model

In the paper [1] we have already considered the approximate model of electron. We note here again that the electron model isn't the basis of our theory, but it allows to understand it better. Here we will make some additions, which make the model correspond better (as we think) to the experimental facts.

## A.1. Parameters of the electron model

We represent the electron as a torus with a radius $r_s = r_p$, where the index "$p$" refers to the circular photon and the index "$s$"

refers to the circular semiphoton. For the radius of the ring cross-section we have taken arbitrarily the same value $r_p$.

Here we suppose that the cross-section torus radius is equal to $r_c$, where $r_c < r_s$ (fig.A1), and that $\frac{r_c}{r_s} = \zeta$, where $\zeta < 1$; then $r_c = r_s \zeta$.

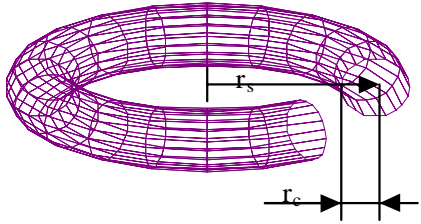

Fig.A1. Electron model

According to the our model the torus ring radius has the value $r_p = \frac{\lambda_p}{2\pi}$, where $\lambda_p$ is the photon's wavelength. The photon characteristics are defined by the electron-positron pair production conditions: the photon energy $\varepsilon_p = 2m_e c^2$ and the circular frequency $\omega_p = \frac{\varepsilon_p}{\hbar} = \frac{2m_e c^2}{\hbar}$. Therefore, the photon wavelength is $\lambda_p = \frac{2\pi c}{\omega_p} = \frac{\pi \hbar}{m_e c}$, and the radius of torus (i.e. of the circular semiphoton) is $r_s = \frac{\hbar}{2m_e c}$.

## A.2. Charge and mass

Now we can calculate the semi-photon model charge (see in detail [1]):

$$q = \frac{1}{\pi} \frac{\omega_s}{c} E_o S_s 2 \int_0^{\frac{\lambda_s}{4}} \cos \, k_s l \, dl = \frac{1}{\pi} E_o S_c, \qquad (A.1)$$

Since $S_s = \pi r_c^2$, we obtain:

$$q = E_o r_c^2 = \zeta^2 E_o r_s^2, \qquad (A.2)$$

The formula (A.2) can be written in the Coulomb's law form as:

$$E_0 = \frac{q}{r_c^2} = \frac{q}{\zeta^2 r_s^2}, \qquad (A.3)$$

Comparing (A.3) with the classical form of Coulomb's law [3]:

$$E = \frac{q}{\varepsilon_0 r}, \qquad (A.4)$$

we result that the $\zeta^2$ corresponds to the vacuum constant $\varepsilon_0$.

The calculation of the mass of the model electron give us the expression [1]:

$$m_s = \frac{\pi \, \zeta^2 E_0^2 r_s^2}{4\omega_s c} \, ,$$ (A.5)

## A.3. Electromagnetic constant

Using equations (A.2) and (A.5) we can write:

$$m_s = \frac{\pi \, q^2}{4\omega_s c r_s^2} \, ,$$ (A.6)

or, taking in account that $\omega_s \cdot r_s = c$ we obtain:

$$r_s = \frac{\pi}{2} \frac{q^2}{2 m_s c^2} \, ,$$ (A.7)

Put here the above values $r_s = \frac{\hbar}{2 m_e c}$, we have:

$$\frac{q^2}{\hbar c} = \frac{2}{\pi} \zeta^2 = \alpha_q \approx 0{,}637 \zeta^2 \, ,$$ (A.8)

Comparing with electromagnetic constant value

$$\frac{e^2}{\hbar c} = \alpha \cong \frac{1}{137} \, ,$$ (A.9)

we obtain

$$\zeta = \sqrt{\frac{\pi}{2} \alpha} = 0{,}107 \, ,$$ (A.10)

Thus, if in electron model we put

$$r_c = \sqrt{\frac{\pi}{2} \alpha} \, r_s = 0{,}107 r_s \, ,$$ (A.11)

then we have the experimental value of the electron charge $e$.

It is not difficult to see that the electromagnetic constant $\alpha$ plays here the role of vacuum constant. Neglecting the geometric coefficient $\frac{\pi}{2}$, we can write (A.3) in the form:

$$E_0 = \frac{q}{\zeta^2 r_s^2} = \frac{q}{\alpha \, r_s^2} \, ,$$ (A.12)

From above we can suppose that the electromagnetic constant has the physical sense of the dielectric constant of the physical vacuum and corresponds to the conversion of the "bare" electron with the radius $r_s$ into the "dressed" electron with classical radius $r_0$.

**A.4 Quantum of magnetic flow of the ring model electron**

From researches of the macroscopic superconductive vortex has been found the universality of the quantum of the magnetic flow through the superconductive ring [10]:

$$\varphi = n\varphi_0 \text{ ,} \tag{A.13}$$

where

$$\varphi_0 = \frac{h}{2e} \text{ ,} \tag{A.15}$$

is the elementary magnetic flow, $h$ is the Plank constant, $e$ is the electron charge, $n = 1,2,3,\ldots$ is integer.

It is not difficult to show that the elementary magnetic flow is also the characteristic of our electron model.

The electron ring is the relativistic current ring. Let's suppose that the current of ring creates the magnetic field $H$, which hold the ring in the equilibrium as the charge particle on the cyclotron orbit. Then for any part of the electron ring the Newton law is right [10]:

$$\Delta e \cdot c \cdot H = \frac{\Delta m \cdot c^2}{r_s} \text{ ,} \tag{A.16}$$

where $\Delta e$ and $\Delta m$ are the parts of charge and mass correspondingly. Integrating (A.21) we obtain:

$$e \cdot c \cdot H = \frac{m_e \cdot c^2}{r_s} \text{ ,} \tag{A.17}$$

According to definition of the magnetic flow:

$$\varphi = \pi \, r_s^2 H \text{ ,} \tag{A.18}$$

Using $r_s$ value and (A.17) we have for the elementary magnetic flow of the electron ring model:

$$\varphi_0 = \frac{h}{4e} \text{ ,} \tag{A.19}$$

The difference between (A.15) and (A.19) may lie in the fact that the superconductive vortex is the 'boson" and the electron is the fermion.

**Conclusion**

As we see, the electromagnetic interpretation of the quantum mechanics is in accordance with modern result of the lepton physics.